\pdfoutput=1
\documentclass[a4page,10pt]{article}
\usepackage[top=0.8in, bottom=0.8in, left=0.5in, right=0.5in]{geometry}
\usepackage{fancyhdr}
\renewcommand{\thispagestyle}[1]{}
\usepackage{times,amsmath,graphicx}
\usepackage[ruled, vlined, boxed]{algorithm2e}
\begin{document}
\title{Incremental Entity Resolution from Linked Documents}

\author{
{Pankaj Malhotra, Puneet Agarwal, Gautam Shroff}%
\vspace{1.6mm}\\
TCS Research\\
Tata Consultancy Services Ltd., Sector 63\\
Noida, Uttar Pradesh, India\\
\fontsize{10}{10}\selectfont\ttfamily\upshape
\{malhotra.pankaj, puneet.a, gautam.shroff\}@tcs.com\\
\vspace{1.2mm}\\
\fontsize{10}{10}\selectfont\rmfamily\itshape
\fontsize{9}{9}\selectfont\ttfamily\upshape
}
\maketitle
\begin{abstract}
\textit{In many government applications we often find that information
about entities, such as persons, are available in disparate data sources such as
passports, driving licences, bank accounts, and income tax records.
Similar scenarios are commonplace in large enterprises having multiple customer, supplier, or partner databases. 
Each data source maintains different aspects of an entity, and resolving entities based on
these attributes is a well-studied problem. However, in many cases documents in one source reference
those in others; e.g., a person may provide his driving-licence number while applying for 
a passport, or vice-versa. These links define relationships between documents of the same entity
(as opposed to inter-entity relationships, which are also often used for resolution).
In this paper we describe an algorithm to cluster documents that
are highly likely to belong to the same entity by exploiting inter-document 
references in addition to attribute similarity. Our technique uses a combination of iterative graph-traversal, locality-sensitive hashing,
iterative match-merge, and graph-clustering to discover unique entities based on a document corpus. A unique feature
of our technique is that new sets of documents can be added incrementally
while having to re-resolve only a small subset of a previously resolved entity-document collection.
We present performance and quality results on two data-sets: a real-world database of companies and a large 
synthetically generated `population' database. We also demonstrate benefit of using inter-document references for clustering 
in the form of
enhanced recall of documents for resolution.}
\end{abstract}

\section{Introduction}
We address a commonly occurring situation where information about entities, such as persons,
are maintained in disparate data sources each describing different aspects
of an entity. Our goal is not only to identify the documents belonging to an entity, but also to merge these documents 
in order to create one document per entity.
For example, information about residents of a country is available in multiple disparate databases like passports, driving licenses, bank accounts, phone-connection
records, and income-tax records. 

We focus on the fact that often there are cases when one type of document of an entity may contain a reference to another
type of document. For example, a passport document may refer to the driving license id of the person as an identity
proof. Such references can be leveraged to identify different documents belonging to the same person, as we describe in
our approach for `entity resolution from linked documents'.

The problem of entity resolution has been addressed in many ways.
Many approaches to entity resolution problem view it as a
problem of data de-duplication \cite{p:duplicateDetectionSurvey}, duplicate detection
\cite{p:adaptiveDuplicateDetection}, or data cleaning \cite{p:dataCleaning}. In these cases the goal is to detect and remove
duplicate records for the same real-world entity, merge-purge \cite{p:mergePurge}, or link records
\cite{p:recLinkageSurvey}. Collectively resolving entities based on relationships between \textit{different} 
entities has been addressed in \cite{p:indrajit:ColBatch}. In contrast, we focus on references
between documents that belong to the \textit{same} entity, and show how this additional information
can be used to enhance the accuracy and performance of entity resolution.

Explicitly comparing all pairs of documents is of quadratic complexity and becomes infeasible for even moderately sized
collections. To achieve scalability `blocking' techniques are generally used \cite{swooshThesis} wherein blocks of documents are obtained such that each block contains a relatively small number of potentially matching documents. Thereafter only documents within each block are exhaustively compered with each other (so the number of comparisons done is greatly reduced) to find matching documents (i.e., documents belonging to same entity). Hence, blocking based entity resolution is usually considerably more efficient than an all-pairs comparison. \textbf{We show how blocking efficacy can be enhanced by exploiting inter-document references}.

In many real-life situations new sets of documents keep arriving incrementally over time. These documents may pertain to new
entities, i.e., the entities that do not exist in the database of resolved entities at the time these documents arrive,
or may pertain to already existing entities. New documents could even potentially alter previously taken decisions about
which sets of documents belong to different entities. Thus the database of resolved entities needs to be appropriately updated
with this new information. \textbf{We show how to update the resolved set of document-entity pairs efficiently, without having
to re-resolve the entire collection.}

The rest of the paper is organized as follows: We begin by formally describing the linked documents scenario in Section
\ref{sec:def}. In Section \ref{sec:algo} we explain the key motivations for our solution and introduce our approach to resolving entities from linked documents, which we call ERLD. In Section \ref{sec:batchER} we the describe the ERLD algorithm in detail and its incremental
variant in Section \ref{sec:IncreMode}. 

In Section \ref{sec:exper} we present performance results on two data-sets: a real-world database of companies, and a large synthetically generated `population'
database. We show how leveraging the inter-document references enhances the quality of the solution by improving recall. We also demonstrate the scalability of ERLD as compared to an pairwise iterative match merge approach \cite{p:swoosh}. Finally we show that resolving a new set of documents against a pre-resolved set of entities can be done efficiently using Incremental ERLD. We conclude in Section \ref{sec:conc} after a brief description of related work in Section \ref{sec:relWork}. 

\section{Linked Documents}\label{sec:def}
We assume that information about real-world entities is available from disparate data sources in the form of documents linked by references, formally defined as follows: \\
\textbf{Document}: A \textit{document} consists of \textit{attributes}, where each \textit{attribute} has a unique name and zero or more values. Each document belongs to a unique real-world \textit{entity}. For example, a passport has attributes \textit{name, father's name, address,} and \textit{date-of-birth}; additionally, one's \textit{driving licence number} may be given as identity proof.\\
\textbf{Match}: Two documents are said to Match if they return \texttt{true} under some \textit{match function}. \\
\textbf{Match function}: A Boolean function defined over two documents that returns \texttt{true} when two \textit{documents} are deemed as belonging to the same entity, and \texttt{false} otherwise. Match functions can be implemented in different ways. For example, they can be based on rules defined over the attribute values of the two documents  being compared: \texttt{true} if name matches AND address matches AND date-of-birth matches, \texttt{false} otherwise. We use such a rule-based Match function in our experiments on one of the data-sets (refer Section \ref{subsec:ResidentsData}). Match functions may also be based on classifiers derived via machine learning: Features for such a classifier can be the scores obtained by comparing the attribute values of the corresponding attributes in the two documents. For example, the Jaccard similarity score can be used to compare name attributes from two documents. Scores from different such similarity measures become features to classify the pairs of documents as \texttt{true} (matching) or \texttt{false} (non-matching) categories, via machine learning. We use a Support Vector Machines (SVM) classifier \cite{svmBurges} in another of our case-studies (refer Section \ref{subsec:CompaniesData}).\\
\textbf{Merge}: A procedure that takes two or more documents as input and produces a new document that has attributes and their values coming from all the input documents. Note that the resulting merged document may have multiple values for some or even all its attributes.\\
\textbf{Entity}: We use the term entity in two senses: (i) real-world entities such as persons and (ii) clusters of documents that are likely to belong to the same real-world entity. Thus, an entity of the type person has (potentially multi-valued) attributes such as name, father's name, address, date-of-birth, etc. The values of 
these attributes are gathered from disparate documents associated with the same entity based on a Match function.

\begin{figure}
	\centering
		\includegraphics[scale=0.25,natwidth=2943,natheight=1569]{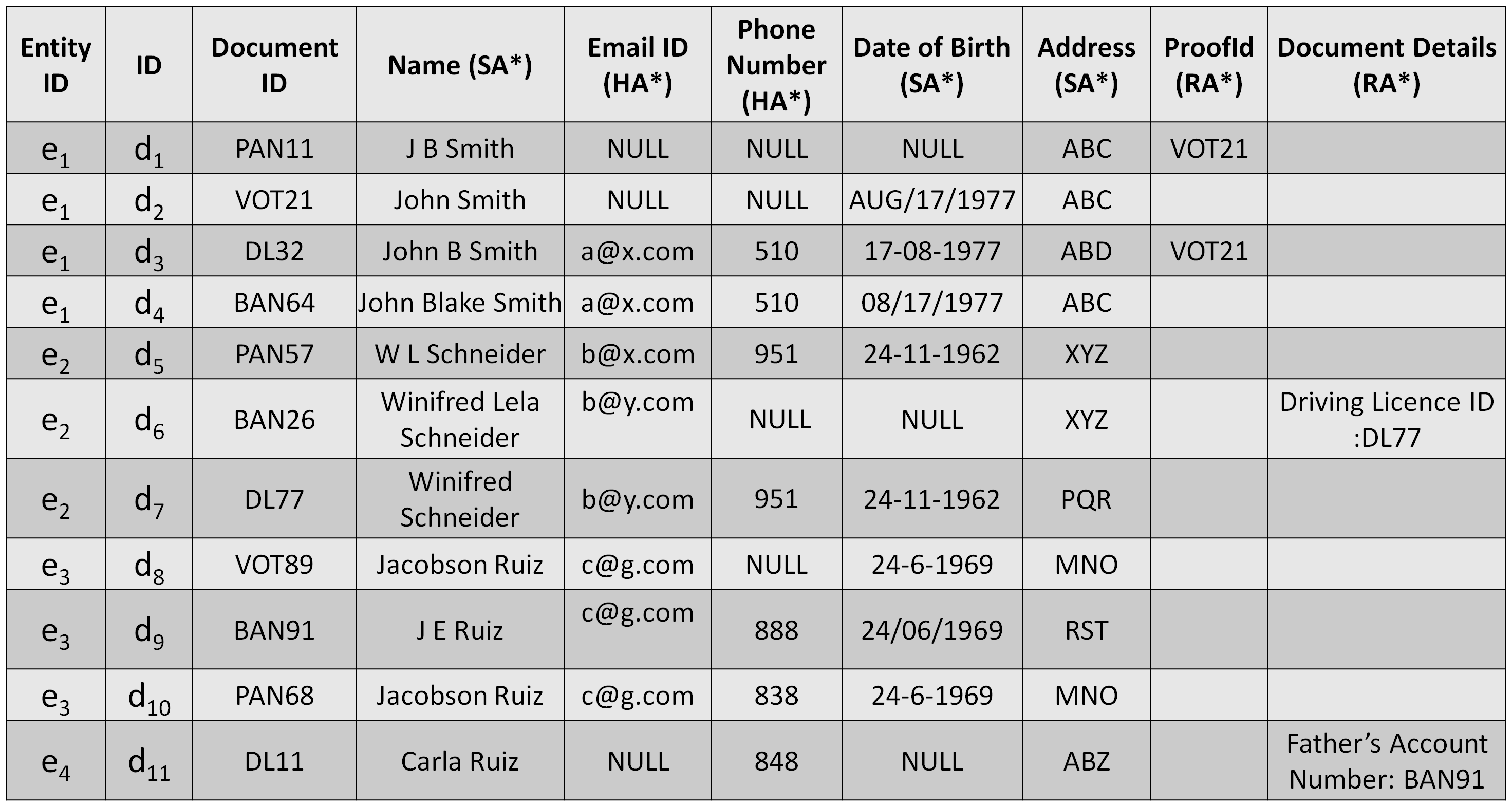}
	\caption{Sample Linked Documents. Here, SA*:Soft Attribute, HA*: Hard Attribute, RA*: Referential Attribute}
	\label{fig:Data}
\end{figure}

Document attributes differ depending what it means for different values of a particular attribute to be considered `matching'; in this context we define \textit{types} of attributes as follows:\\
\textbf{Soft Attribute}: An attribute for which two values may be considered as matching under some matching criterion, even if they are not textually same/equal. For example, different variations of a person's \textit{name} (as shown in Figure \ref{fig:Data}) can be considered to be matching even if they are not textually identical.\\
\textbf{Hard Attribute}: An attribute for which two values are considered to be matching only if they are textually identical: For example, \textit{Phone number} and \textit{email-id} are hard attributes in Figure \ref{fig:Data}.\\
\textbf{Unique Attribute}: An attribute that has a unique value for each real-world entity. Clearly a unique attribute must also be hard, i.e, two values are considered to be matching only if they are textually identical. For a real-world entity possessing multiple documents, the value of any particular unique attribute must be the same in all of them (apart from cases of deliberate obfuscation or data errors). Also, two entities cannot have the same value for a unique attribute. For example, an entity can have only one \textit{passport number}, and no two entities can have same \textit{passport number}. Unique hard attributes are potential primary keys of a document. The idea of unique attributes has been defined in a similar way in \cite{constraints}, where an attribute has a hard uniqueness constraint if it can take a unique value or no value for each real-world entity.

Note that an entity \textit{can} have multiple values for a soft or hard attribute, but not for a unique attribute. For example, a person can have multiple names and phone numbers but not multiple passport numbers. Further, different entities can have same value for a soft attribute or a hard attribute but not for a unique attribute. For example, two people can have same name (soft attribute) and can share same phone number (hard attribute) but not same passport number (unique attribute). On the other hand, violations of the uniqueness property for values of a unique attribute can be an indication of deliberate obfuscation, e.g., a person wanting to hide under multiple identities, or of data errors in a poor quality database. The implications of uniqueness and how it can exploited may differ, e.g., for entity resolution or alternatively for detecting fraud or other suspicious behaviour.\\
\textbf{Referential Attribute}: An attribute of a document that contains the value of a hard or unique attribute of \textit{another} document. For example, an attribute named \textit{Driving Licence ID} in a passport document can be considered as a referential attribute. A hard attribute can also be a referential attribute. For example, the \textit{phone number} (a hard-attribute) of a person in a bank-account document may be the primary-key of the phone-connection document for that person. The \textit{Proof ID} and \textit{Document Details} attributes in Figure \ref{fig:Data} contain inter-document references, and can be considered to be referential attributes.

We classify \textit{referential attributes} into two types: (1) Explicit Referential Attributes, (2) Implicit Referential Attributes:
The value of an explicit referential attribute exactly the value of a hard or unique attribute. On the other hand, for an implicit referential attribute,
a \textit{part} of its value \textit{contains} a hard or unique attribute of another document. It is unknown which part of the value of an implicit referential attribute is the actual reference. For example, if a passport document has an attribute named \textit{Driving Licence ID} with value equal to ``DL123", then the passport document  makes an explicit reference to the \textit{driving licence} document with primary key ``DL123", whereas if the value of the `description' field in the passport document is ``Applicant's DL\# DL123", then the reference to the driving license document is implicit. For the example in Figure \ref{fig:Data}, \textit{Proof ID} is an explicit referential attribute whereas \textit{Document Details} can be an implicit referential attribute, when, for example it \textit{contains} the value ``BAN91'' as part of its textual content.

A document referring to another document is likely to indicate one of the two things: (1) documents belong to the same entity, (2) documents belong to different entities which are related to each other. For example, a description field in a bank-account document containing a mobile number may imply that the mobile number referred is of the same person or that of a relative (such as a parent) of the person. If this mobile number is also present in another document, then the two documents potentially belong to the same person or two persons who are related to each other. However in this paper we do not consider or exploit \textit{explicit} references between documents belonging to different entities. Further \textit{implicit} references to other entities are treated as noise for our purposes.
\section{Entity Resolution from Linked Documents}\label{sec:algo}
\begin{figure*}
	\centering
		\includegraphics[width=\textwidth,natwidth=2806,natheight=651]{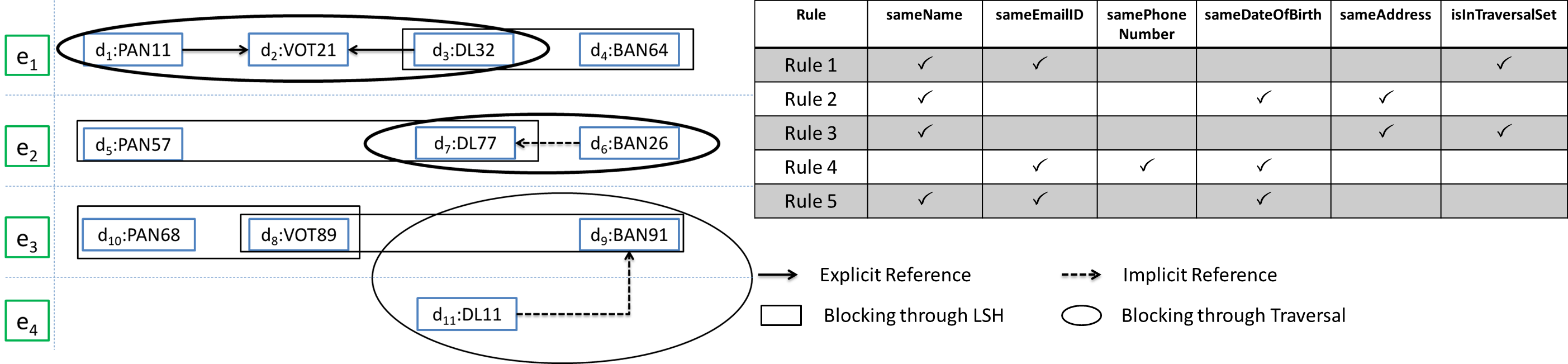}
	\caption{A pictorial view of Sample Linked Documents of Figure \ref{fig:Data} with Rules for Matching}
	\label{fig:SampleData}
\end{figure*}
\textbf{Entity Resolution (ER)}: Given a collection of \textit{documents} D = \{d$_{1}$, d$_{2}$,..., d$_{N}$\}, where each \textit{document} d$_{i}$ belongs to a unique \textit{entity}, determine a collection of entities E = \{e$_{1}$, e$_{2}$,..., e$_{K}$\}  where each e$_{i}$ is obtained by applying the \textit{Merge} function on a subset of document(s) coming from D that \textit{Match} under a given \textit{Match} function. For example, different of documents belonging to the same person, such as Voter-Card, Passport, and Driving-Licence, will get merged to form one entity.

\subsection{Overview of ERLD Algorithm}
The goal of our \textbf{E}ntity \textbf{R}esolution from \textbf{L}inked \textbf{D}ocuments algorithm (ERLD) is to group documents into entities so that all the documents in an entity correspond to the same real-world entity. When the number of documents that need to be resolved is large, comparing every document with every other document to group the matching documents is infeasible. Using blocking techniques \cite{indexingSurvey, swooshThesis} it is possible to avoid comparisons between documents that are highly unlikely to belong to the same entity. To avoid unnecessary comparisons we exploit textual similarity as well as inter-document references to form blocks/clusters of documents that are highly likely to belong to the same entity. Once such clusters are formed, we resolve the entities in each cluster using R-Swoosh based Iterative Match-Merge \cite{p:swoosh}. Finally we consolidate the results from each cluster using graph-clustering, i.e.,  by finding connected components in a graph of documents with edges defined using the partial-entity assignments (see Section \ref{subsec:CC}). 
\begin{figure}
	\centering
		\includegraphics[scale=0.4,natwidth=1461,natheight=766]{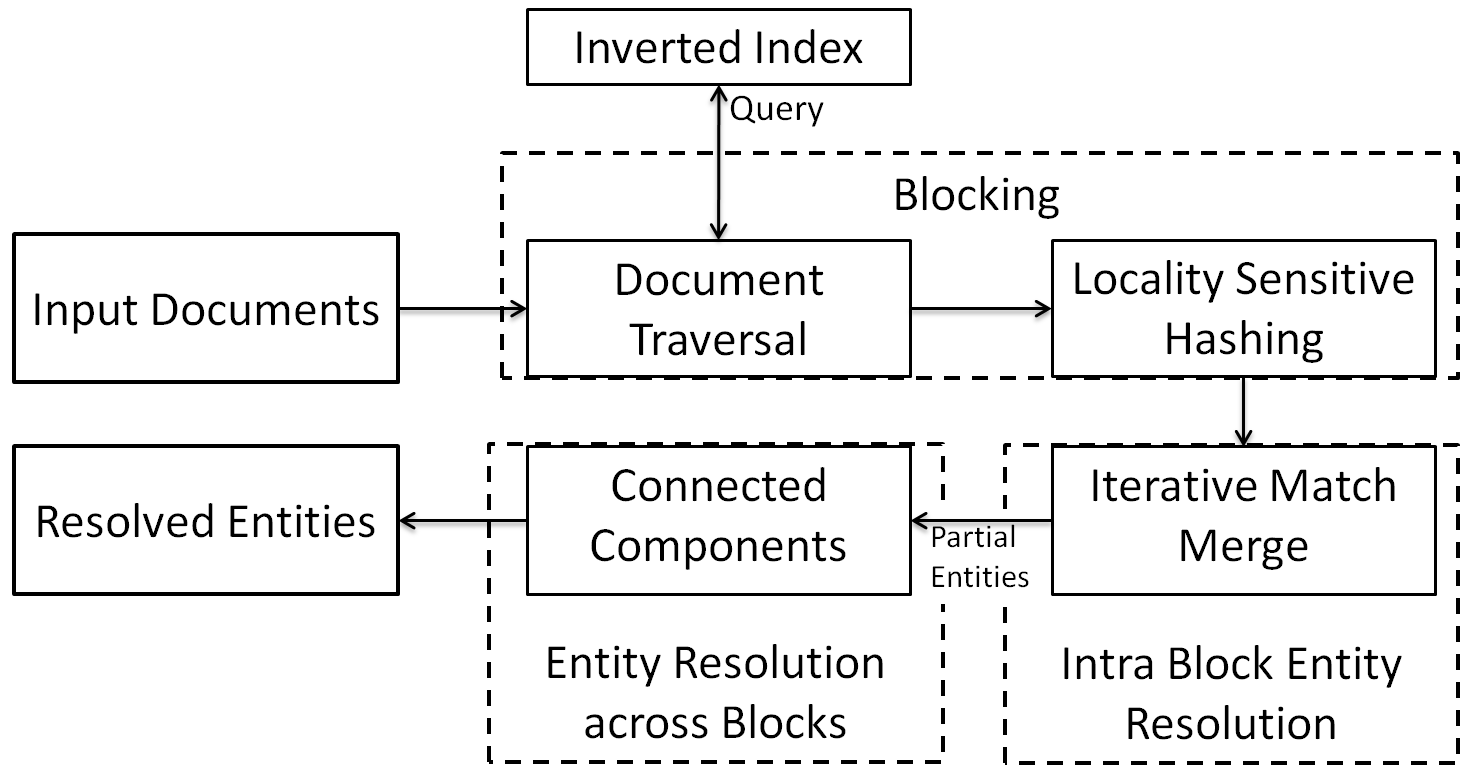}
	\caption{Entity Resolution from Linked Documents (ERLD)}
	\label{fig:EntResProcess}
\end{figure}
\subsection{Motivating Example}
Consider the 11 documents in the collection D = \{d$_{1}$, d$_{2}$,..., d$_{11}$\} containing information about 4 entities, as depicted in Figure \ref{fig:Data} (and later used in Figures \ref{fig:SampleData}, \ref{fig:FlowAfterDT-LSH1}, and \ref{fig:BucketResCC}).  The documents d$_{1}$ to d$_{4}$ belong to entity e$_{1}$, d$_{5}$ to d$_{7}$ belong to e$_{2}$, d$_{8}$ to d$_{10}$ belong to e$_{3}$, and d$_{11}$ belongs to e$_{4}$. Given D and a \textit{Match} function, the goal is to determine E = \{e$_{1}$, e$_{2}$, e$_{3}$, e$_{4}$\} such that e$_{1}$ = Merge(d$_{1}$, d$_{2} $, d$_{3}$, d$_{4}$), e$_{2}$ = Merge(d$_{5}$, d$_{6}$, d$_{7}$), e$_{3}$ = Merge(d$_{8}$, d$_{9}$, d$_{10}$), and e$_{4}$ = d$_{11}$. The documents are of 4 types obtained from 4 different sources: Income Tax documents (PAN\footnote{Income-tax `permanent account number' in the Indian context}), Voter-Card (VOT), Driving-Licence (DL), and Bank Account Details (BAN). The attributes of the documents are \textit{Document ID, Name, Email ID, Phone Number, Date of Birth, Address, Proof ID}, and \textit{Document Details}. Documents d$_{1}$ and d$_{3}$ make explicit reference to d$_{2}$ through the \textit{Proof ID} attribute. Documents d$_{7}$ and d$_{11}$ make implicit reference to d$_{6}$ and d$_{9}$ respectively through the \textit{Document Details} field. Documents d$_{3}$ and d$_{4}$ have high textual similarity, i.e., the values of their corresponding attributes are very similar. Same is the case with documents d$_{5}$ and d$_{6}$, documents d$_{8}$ and d$_{9}$, and documents d$_{8}$ and d$_{10}$. 

Note that we have assumed that the schemas of different document types have already been matched, i.e., we already know which fields in each document refer to a person's Name. When documents are completely unprocessed, even this can be a difficult problem; however, techniques for schema mapping are already available, such as \cite{cupidSchemaMatching}.

The Match function is based on a disjunction of rules shown in Figure \ref{fig:SampleData}. A rule is satisfied when the Boolean functions based on the tick-marked columns are all \texttt{true}. For example, for two documents \textit{r }and \textit{r'}, Rule 2 reads: \emph{sameName}(\textit{r},\textit{r'}) \texttt{AND} \emph{sameDateOfBirth}(\textit{r},\textit{r'}) \texttt{AND} \emph{sameAddress}(\textit{r},\textit{r'}). Here, \emph{sameName}(\textit{r},\textit{r'}), \emph{sameDateOfBirth}(\textit{r},\textit{r'}) and \emph{sameAddress}(\textit{r},\textit{r'}) are Boolean functions. Two documents \textit{r} and \textit{r'} match based on Rule 2 when this clause is satisfied, and two documents ``match'' if at least one of the five rules is satisfied. 

Note that each of the Boolean functions such as \mbox{\emph{sameName}()} may be implemented by checking for exact textual match or a approximate match using some similarity measure. Now we briefly describe the major steps our \textbf{ERLD} algorithm using the above motivating example.

\subsection{Major steps of ERLD}
\begin{enumerate}
	\item \textbf{DT: Document Traversal to exploit links between documents}: If a document makes a reference to some other document, then the two documents may belong to the same real-world entity. For example, a person may use her \textit{driving license number} as a proof of her name/identity to get her passport, or a blog profile of a person may link to her twitter account. Such references help to identify some of the documents belonging to the same entity which in turn help to resolve it. This is particularly helpful when two documents belonging to the same entity do not have sufficient attributes-based similarity.	

For example, a person may use her maiden name in her driving license, which in turn is used as identity proof (together with a marriage certificate) while applying for a passport. The driving license and passport may not match purely based on attribute similarity, but the inter-document reference forces them to be assigned to the same entity.

During Document Traversal we traverse the links between documents as defined by \textit{referential attributes}. For example, documents d$_{1}$, d$_{2}$, and d$_{3}$ in Figure \ref{fig:SampleData} are linked to each other. Such links can be captured through DT, and based on rules Rule 1 and Rule 3, which explicitly include a check for two documents being related by references (the \mbox{\emph{IsInTraversalSet}()} function, which is true if two documents are connected by a path of references), it can be established that the documents belong to the same entity. We discuss Document Traversal in detail in Section \ref{subsec:DT}.\\
	\item \textbf{LSH: Similarity-based hashing to exploit textual similarity between documents}: Documents with similar content are likely to belong to the same real-world entity. For example, if the values of \textit{name}, \textit{address}, and \textit{phone number} attributes are same in two documents (as is the case with documents d$_{3}$ and d$_{4}$ in Figure \ref{fig:SampleData}), it is very likely that the documents belong to the same real-world entity. On the other hand, if only \textit{name} matches, and \textit{address} and \textit{date-of-birth} differ, the two documents are less likely to belong to the same entity. To avoid unnecessary comparisons between documents with very low textual similarity, we use \textit{MinHash} based \textit{Locality Sensitive Hashing} \cite{minhash, lsh} to cluster similar documents, and perform comparisons only amongst documents belonging to the same cluster.  As a result, after DT and LSH, documents that are not textually similar and are not linked to each other through references are unlikely to be even compared. We describe how clusters are computed using LSH in Section \ref{subsec:LSH} .
	\item \textbf{IMM: Iterative Match-Merge to handle incomplete information in documents}: A document obtained by merging two matching documents may contain more information about the entity than the information revealed by either of the two documents when they are considered separately. To exploit this, we use \textit{Iterative Match Merge} (refer to Section \ref{subsec:IMM}) based on the R-Swoosh algorithm described in \cite{p:swoosh}. 
For the example in Figure \ref{fig:SampleData}, documents d$_{5}$ and d$_{6}$ do not satisfy any of the rules (they only satisfy \emph{sameName}(d$_{5}$,d$_{6}$) and \emph{sameAddress}(d$_{5}$,d$_{6}$)), and hence don't match. Similary d$_{5}$ and d$_{7}$ don't match (they satisfy \emph{sameName}(d$_{5}$,d$_{6}$), \emph{samePhoneNumber}(d$_{5}$,d$_{6}$), and \emph{sameDateOfBirth}(d$_{5}$,d$_{6}$)). But d$_{6}$ and d$_{7}$ (linked through DT) match based on Rule 1. When these two matching documents are merged to get a new document (say, d$_{67}$), we now know the values of five attributes (\textit{name, email-id, phone number, date-of-birth}, and \textit{address}) for d$_{67}$. We have additional information due to the values of \textit{phone number} and \textit{date-of-birth} for an entity coming from document d$_{7}$, and two different addresses for the same entity coming from d$_{6}$ and d$_{7}$. This additional information can be useful in future matching operations. The merged document d$_{67}$ has sufficient information to match with d$_{5}$ based on Rule 2. (Note that d$_{5}$ and d$_{67}$ are considered for comparison because they are in same block due to the combined effect of LSH and DT. In other words, LSH puts d$_5$ and d$_6$ in same block while DT links d$_6$ to d$_7$. So all the three documents end up being in the same block. The details of this process are described in Section \ref{sec:batchER}.) 
	\item \textbf{CC: Connected Components} The combination of DT and LSH results in blocks of documents that are resolved into entities. Still, the possibility remains for documents belonging to a single entity to get mapped to more than one block. For example, one LSH block due to \textit{Name}, and another due to \textit{Address}. Yet the two `partial entities' formed in each such block are still connected by the fact that they both contain overlapping documents. Therefore, we finally consolidate \textit{partial entities} emerging from all the blocks by computing connected components in a graph of documents where an edge exists between two documents if they belong to the same partial entity. If a pair of partial entities e$_a$ and e$_b$ happen to share at least one of the original documents d$_i$, they end up being in the same connected component and all the documents in them get merged.
\end{enumerate}

\textit{Summary:} In a scenario where a link between two documents means that both the documents belong to the same entity, and all documents belonging to the same entity are also  linked to each other, then the entity can be resolved only on the basis of DT. However, when the documents belonging to the entity form more than one component in the graph of document-linkages, we need other ways to connect all the documents belonging to the same entity. For such cases, a blocking-scheme based on a combination of DT and LSH can help, where DT helps to discover links based on referential attributes and LSH helps to discover potentially matching documents based on non-referential attributes. For the example in Figure \ref{fig:SampleData}, d$_{3}$ and d$_{4}$ get hashed to the same block on the basis of LSH. Documents d$_1$ and d$_2$ do not have high textual similarity with either d$_3$ or d$_4$, and hence using only LSH for blocking, no single block would form containing d$_1$, d$_2$, d$_3$, and d$_4$, and the entity e$_1$ would not get completely resolved. It is because of DT that d$_1$ and d$_2$ get linked to d$_3$. The blocking scheme based on combination of DT and LSH leads to at least one such block where all the four documents belonging to e$_1$ co-exist, and then using IMM on all the four documents results in the desired entity e$_1$.

We consider two different scenarios in which Entity Resolution is encountered in practice: 1) \textit{Batch mode}, when an entire collection of documents need to be resolved within themselves, and 2) \textit{Incremental mode}, when a new, relatively small collection of documents need to be added to an already resolved entity-document collection. We first describe in detail the batch mode process (ERLD) and then describe how the batch mode solution can be modified to perform Incremental Entity Resolution (Incremental ERLD).
\section{ERLD Algorithm Details}\label{sec:batchER}
We now describe the steps of the ERLD algorithm (see Figure \ref{fig:EntResProcess}
 and Algorithm \ref{algo:BatchAlgo}) in the order they are performed.

\begin{algorithm}\label{algo:BatchAlgo}
 \SetAlgoLined
 \LinesNumbered
 \SetKw{KwTo}{in}
 \textbf{Input: }\textit{D = \{d$_{1}$,d$_{2}$,...,d$_{N}$\}, raw documents.}\\
 \textbf{Parameters: }\textit{m = \#Hash functions required to get a bucket id; n  = \#BucketIds generated per
document.}\\
 \textbf{Output: }\textit{E = \{e$_{1}$,e$_{2}$,...,e$_{K}$\}, resolved entities.}\\
 \BlankLine
 \tcp*[h]{key is bucket-id, value is list of documents}\;
 $buckets \leftarrow$ empty hash table\; 
 $bucketIds \leftarrow$ empty array\;
 \BlankLine
  Build inverted index\;
 \BlankLine
 \For{$i\leftarrow 1$ $to$ $N$}{
	$d_{i}$.traversalSet $\leftarrow$ getTraversalSet(d$_{i}$)\;\label{algo:travSet}
	$d_{i}$.minhashSignature $\leftarrow$ getMinhashSignature(d$_{i}$)\;
	$bucketIds$ $\leftarrow$ getBucketIds(d$_{i}$.minhashSignature)\;
	\For{$j\leftarrow 1$ $to$ size($bucketIds$)}{\label{algo:lshStart}
	 $bucket \leftarrow$ empty list\;
	 $bucket$ $\leftarrow$ $buckets$.get($bucketIds$[j])\;
	 $bucket$.add(d$_{i}$)\;
	 $bucket$.add(d$_{i}$.traversalSet)\;
	 $buckets$.put($bucketIds$[j], $bucket$)\;\label{algo:lshEnd}
	}
	}
	\BlankLine
	$docs \leftarrow$ empty list\;
	$partialEnts \leftarrow $empty list\;
	\ForEach{$key$ $\KwTo$ $buckets$.keySet}{
	$docs \leftarrow buckets$.get($key$)\;
	$partialEnts\leftarrow$ RSwoosh($docs$)\; \label{algo:RSwoosh}
	$edges \leftarrow$ generateEdgeList($partialEnts$)\;\label{algo:edgeList}
	$edgeList$.add($edges$)\;
	}
	\BlankLine
	\tcp*[h]{key is component-id, value is list of documents}\\
	$conComps\leftarrow$ empty hash table\;
	$conComps \leftarrow$ ConnectedComponents($edgeList$)\;
	\BlankLine
	\textbf{int} $i \leftarrow 1$\;
	\ForEach{$component$ $\KwTo$ $conComps$} {
	$docs \leftarrow component$.getDocs()\;
	$e_{i} \leftarrow$ merge($docs$)\;\label{algo:mergeCC}
	$i \leftarrow i$+1\;
	}
	\BlankLine
 \caption{ERLD Algorithm}
\end{algorithm}
\begin{figure*}
	\centering
		\includegraphics[scale=0.5,natwidth=2228,natheight=681]{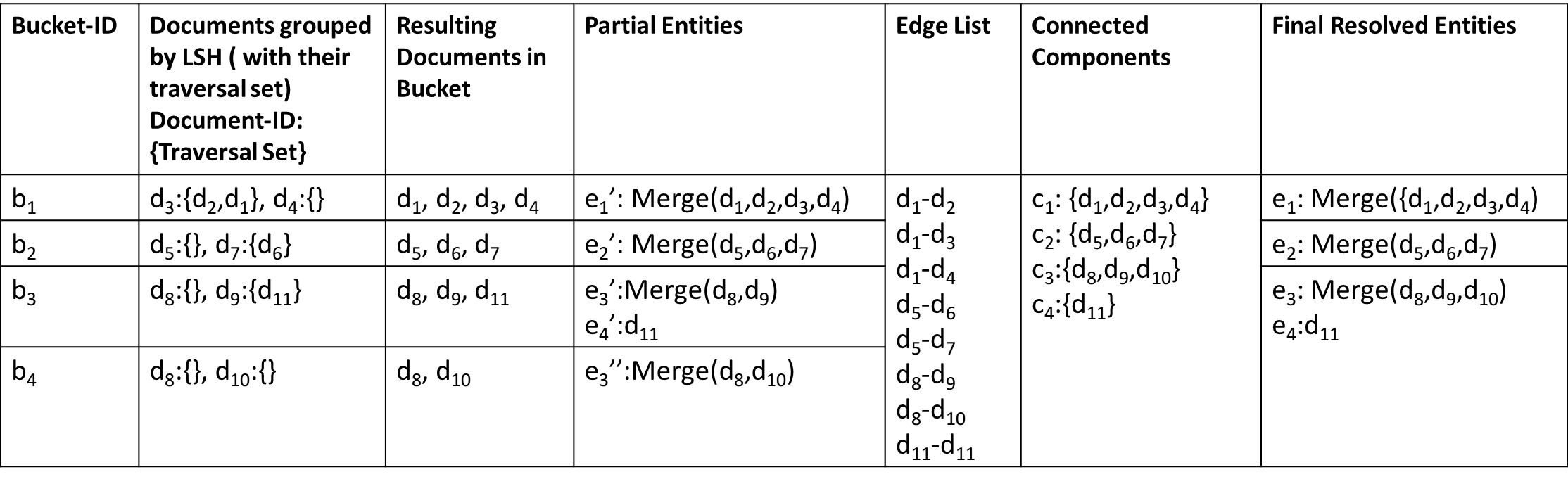}
	\caption{Example of ERLD Flow after DT and LSH have been done for documents in Figure \ref{fig:Data}}
	\label{fig:FlowAfterDT-LSH1}
\end{figure*}
\subsection{Document Traversal}\label{subsec:DT}
Documents are indexed by a primary-key-based database, where a concatenation of the document type
and a unique attribute for that document type is used as primary key. For example, PAN11 indicating document with unique PAN number 11 in the set of PAN documents.
Additionally, an inverted index is built over all the \textit{referential attributes} of the documents so that these attributes become searchable.

Consider the documents as nodes in a graph and the references between documents as directed edges in the graph. The
direction of an edge is determined as follows: if document d$_{i}$ refers to document d$_{j}$ either
explicitly or implicitly, then the edge is directed from d$_{i}$ (the source document) to d$_{j}$ (the
referred document). Going downwards in the graph (starting from a node) means traversing the graph along the direction
of the edges, i.e., from the source documents towards the referred documents, whereas, going upwards means traversing in
the direction opposite to the edge-direction, i.e., from the referred document to the source document.

We maintain a \textit{traversal set} for a document, which is a set of documents connected to it via some path or paths. The
\textit{traversal-set} for a document is generated using a combination of Downstream Traversal (DST) and Upstream Traversal (UST), which we describe next:

\textbf{\textit{Downstream Traversal (DST)}}: Starting from a document, DST captures all the documents reachable from it
going in the downward direction, considering only the edges due to explicit references. The edges due to the
explicit references made by a document are captured by searching the values of the \textit{explicit referential
attributes} in the document via the primary key of the database containing all the documents. 

All the documents obtained by DST on a given document form the
\textit{downstream traversal set} for it. For the example in Figure \ref{fig:SampleData}, 
considering only the explicit references, the downstream traversal set for d$_{1}$ is \{d$_{2}$\}, and for d$_{3}$ is \{d$_{2}$\}. The downstream traversal set for rest of the documents is $\emptyset$.
As another example, considering only the explicit references, suppose \textit{r$_{1}$} refers to \textit{r$_{2}$},
\textit{r$_{2}$} refers to \textit{r$_{3}$}, and \textit{r$_{3}$} refers to \textit{r$_{4}$} and \textit{r$_{5}$}. Then,
the downstream traversal set for \textit{r$_{1}$} is \{\textit{r$_{2}$}, \textit{r$_{3}$}, \textit{r$_{4}$},
\textit{r$_{5}$}\}. The downstream traversal set for \textit{r$_{2}$} is \{\textit{r$_{3}$}, \textit{r$_{4}$},
\textit{r$_{5}$}\}, and for \textit{r$_{3}$} is \{\textit{r$_{4}$}, \textit{r$_{5}$}\}. The downstream traversal set for
\textit{r$_{4}$} and  \textit{r$_{5}$} is $\emptyset$.

Note however, in DST we do not capture the documents that are implicitly referred in a document because we do not know beforehand which
part of the value of the implicit referential attribute should be searched for using the primary key of the document database. For the example in Figure \ref{fig:Data}, for d$_{6}$, the value of the implicit referential attribute \textit{Document Details} is ``Driving License ID:DL77''. Here, we do not know beforehand that ``DL77'' is the key which should be searched for. Implicit references are captured in a different manner through UST, exploiting our inverted index, which we discuss next.

\textbf{\textit{Upstream Traversal (UST)}}: Starting from a document, UST captures all the documents reachable from it
going in the upward direction, considering the edges due to both explicit and implicit references. Apart from the
direction of traversal, UST is different from DST because it also captures the implicit references. For doing UST, an
inverted index over the explicit and implicit referential attributes of all the documents is required. In UST,
the primary key and the values of the explicit referential attributes for the given starting document are
searched using the inverted index. The documents retrieved through this search are put in the \textit{upstream traversal
set} for the document. 

For example, consider a situation where \textit{r$_{1}$} makes an implicit reference to
\textit{r$_{2}$}, which in turn makes implicit references to \textit{r$_{3}$} and \textit{r$_{4}$}. Also, assume that
\textit{r$_{1}$} makes an explicit reference to \textit{r$_{5}$}. Then the upstream 
traversal set after a single step of UST for \textit{r$_{2}$} will be \{\textit{r$_{1}$}\}, and for both
\textit{r$_{3}$} and \textit{r$_{4}$}, the upstream traversal set will be \{\textit{r$_{2}$}\}. The upstream traversal
set for \textit{r$_{5}$} will be \{\textit{r$_{1}$}\}, and the upstream traversal set for \textit{r$_{1}$}  will be
$\emptyset$. For the example in Figure \ref{fig:SampleData}, the upstream traversal set for d$_7$ is \{d$_6$\}, for
d$_9$ is \{d$_{11}$\}, for d$_2$ is \{d$_1$, d$_3$\}. The upstream traversal set for the rest of the documents is
$\emptyset$.

For each document, DST and UST are used to get a traversal set (see line \ref{algo:travSet} in Algorithm
\ref{algo:BatchAlgo}): First, a single step of DST is done for the starting document, and the downstream
traversal set thus obtained is added to the traversal set. This is followed by a single step of UST for all the
documents in the traversal set and the starting document itself. The documents retrieved from this UST step are
used to augment the traversal set. This process of single step of DST followed by a single step of UST is one
\textit{DST-UST step}. This DST-UST step can be repeated a number of times on the documents that get added after each
DST-UST step to the \textit{traversal set}, with a constraint on the maximum number of DST-UST steps and/or the maximum
number of documents retrieved after a single step of UST for a document. 

If the number of documents retrieved in a
single step of UST for a document is more than a threshold, then the retrieved 
documents are not added to the traversal set. This constraint is particularly important in some cases. For example, a
\textit{homepage-url} of a company may appear as an implicit reference in documents belonging to many of its
employees. If this \textit{url} also appears as a value of some explicit referential attribute in some document
in the database, then UST for this document will retrieve a large number of documents. Most of these documents are very
unlikely to belong to the same entity to which the starting document belongs, and should be discarded as noisy results. The constraints
can sometimes come from the domain knowledge. For example, if it is known that there can be a maximum of 5 types of
documents for an entity and one document of the entity can only refer to another document for the same entity, then the
maximum DST-UST steps required to capture all the documents connected to any of the documents is 4. (This happens when
the graph is such that all the 5 nodes are connected to each other as 
r$_{1}$-r$_{2}$, r$_{2}$-r$_{3}$, r$_{3}$-r$_{4}$, r$_{4}$-r$_{5}$, i.e. r$_{1}$ refers to r$_{2}$, r$_{2}$ refers to
r$_{3}$, and so on.)

After traversal, any document has a traversal set associated with it, which is a set of documents potentially
belonging to the same entity as the document. To get further potential matches for any document and avoid unnecessary
comparisons, we use Locality Sensitive Hashing (LSH). 

\subsection{Locality Sensitive Hashing}\label{subsec:LSH}
Consider a set of words created from the values of the attributes (except the referential attributes) in a document. If
two documents have a large number of words in common, i.e., the Jaccard similarity coefficient of the corresponding sets of words for two documents is high, they should be considered for comparison as compared to two documents which do not have many words in common.

LSH hashes the documents to \textit{buckets} such that the probability that two documents get hashed to the same bucket is equal to the Jaccard similarity coefficient of the set of words corresponding to the two documents \cite{p:massiveDM}. The buckets are formed such that documents with high Jaccard similarity  are more likely to end up in the same bucket. Note: The traversal set of a document is also put in the buckets the document belongs to, so as to retain blocking information from the previous DT stage.

Each bucket is a key-value pair, where `key' is the bucket-id, and value is the set of documents which get hashed to
this `key' along with their traversal sets. So, after hashing has been done for each document, each bucket contains
documents which either have high textual similarity, or share explicit and/or implicit references (see
lines \ref{algo:lshStart} to \ref{algo:lshEnd} in Algorithm \ref{algo:BatchAlgo}). The documents which belong to the
same bucket are considered for Iterative Match-Merge, which we discuss in Section \ref{subsec:IMM}.

We use standard LSH via minhashing:  Each word occurring in any of the documents is mapped to a unique integer. For each
document, we consider the set of integers $s$ corresponding to the set of words it contains. The minhash for the set $s$
is calculated as the $min$($(ax_{i}+b)$ mod $p$), $\forall x_{i}\in s$ (where $a$ and $b$ are random integers, and $p$
is a large prime). Different combinations of $a$ and $b$ determine different hash functions (see \cite{minhash} for
details). The probability that two documents, d$_{i}$ and d$_{j}$ (each containing sets of words w$_{i}$ and w$_{j}$
respectively) have the same minhash values for a minhash function is equal to the Jaccard similarity coefficient,
J(w$_{i}$, w$_{j}$). 

For each document, we generate $m \times n$ hash values using $m \times n$ minhash functions. From
these $m \times n$ hash values, $m$ of the hash values are combined together using string concatenation to get a single
\textit{bucket-id}, resulting in a total of $n$ such bucket-ids for each 
document. Due to this concatenation to get bucket-ids, the probability that two documents, d$_{i}$ and d$_{j}$, have at
least one same bucket-id is equal to $1-(1-(J(w_{i}, w_{j}))^{m})^{n}$. 

These two steps, i.e., concatenation and 'or-ing' of hash functions, reduce the probability that
two documents, d$_{i}$ and d$_{j}$, with very low Jaccard similarity (close to 0) will get hashed to the same bucket,
compared to the earlier probability of \mbox{J(w$_{i}$, w$_{j}$)}. Similarly, the probability that two documents
d$_{i}$ and d$_{j}$ with very high Jaccard similarity (close to 1) will get hashed to the same bucket is increased. 

\begin{figure}
	\centering
		\includegraphics[scale=0.4,natwidth=1501,natheight=781]{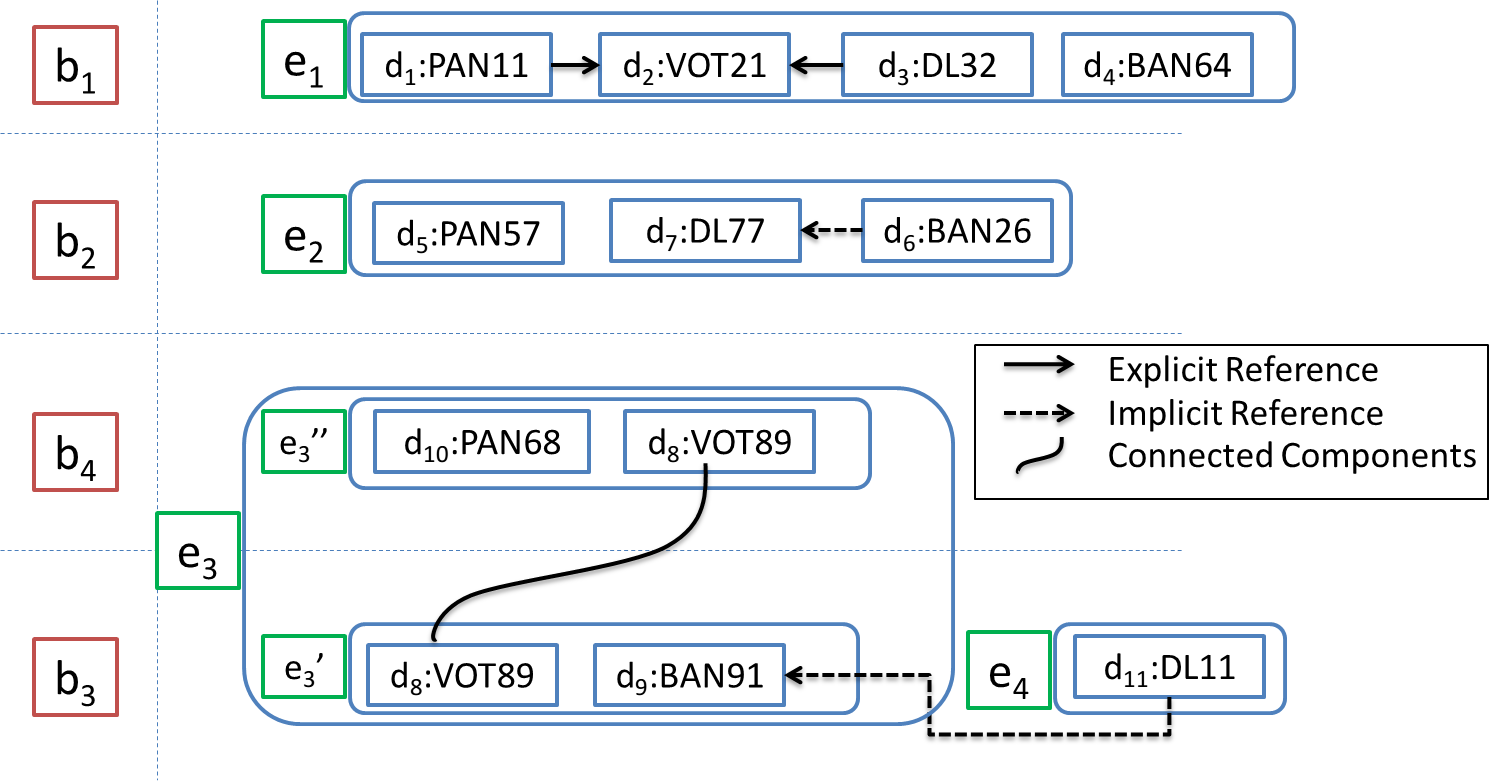}
	\caption{Sample buckets for the example in Figure \ref{fig:Data}}
	\label{fig:BucketResCC}
\end{figure}

For the example in Figure \ref{fig:SampleData}, documents d$_5$ and d$_6$ have high textual similarity and end up being
in at least one same bucket (bucket b$_{2}$ in Figure \ref{fig:FlowAfterDT-LSH1}). Also, since the traversal set for d$_7$ is \{d$_6$\}, the document d$_6$ also comes in the same bucket along
with d$_7$. So, the blocking based on DT followed by LSH helps in resolving the entity e$_2$ (refer Figure \ref{fig:FlowAfterDT-LSH1} and \ref{fig:BucketResCC}).

The documents in a bucket (cluster) go through the Iterative Match-Merge (IMM) process using R-Swoosh \cite{p:swoosh},
which we discuss next.
\subsection{Iterative Match-Merge}\label{subsec:IMM}
We use R-Swoosh, an iterative match-merge process to resolve the documents in each bucket, using the given
\textit{Match} function and the \textit{Merge} procedure (see line \ref{algo:RSwoosh} in Algorithm \ref{algo:BatchAlgo}). We briefly
introduce R-Swoosh as described in \cite{p:swoosh}, and implemented in
SERF\footnote{http://infolab.stanford.edu/serf/\#soft}. Consider two sets of documents, \textit{I} and \textit{I$'$}. To
begin with, set \textit{I} contains all the documents from a bucket, and set \textit{I$'$} is empty. \textit{I$'$}
contains the documents that have all been compared with each other. R-Swoosh iterates over the documents in set
\textit{I}. Iteratively, a document \textit{d} is  removed from \textit{I}, and compared to every document in
\textit{I$'$}. If no matching document is found for \textit{d} in \textit{I$'$}, it is added to \textit{I$'$}. On the
other hand, if a document \textit{d$'$} is found in \textit{I$'$}, such that \textit{d} matches \textit{d$'$}, then
\textit{d$'$} is removed from 
\textit{I$'$}, and a new document obtained through \textit{Merge}(\textit{d}, \textit{d$'$}) is added to \textit{I}.
Although, \textit{d$'$} does not match any other document in \textit{I$'$}, the new merged document may match some
document in  \textit{I$'$}. The new document is added to \textit{I}, so that it is compared with every other document
either in \textit{I} or \textit{I$'$}. In the end, the set \textit{I} is empty, and \textit{I$'$} contains the final
result of R-Swoosh on the documents in the bucket. 

It is to be noted that we do not compare a pair of documents twice,
and rather maintain two sets: one of \textit{matching pairs} and another of \textit{non-matching pairs}. If a pair of documents (which has already been compared in a bucket) is encountered again (in another bucket), the Match function is not computed again. If a pair of documents exists in the set of matching pairs, the value of Match function is taken to be \texttt{true}; if it exists in the set of non-matching pairs, the value is taken to be \texttt{false}; else the Match function is computed for that pair, and based on the value of the Match function, either the set of matching pairs or that of non-matching pairs is updated.

The documents for one real-world entity can co-occur in multiple buckets. So, same entity can be obtained (as a result
of \textbf{IMM} process) from multiple buckets. Also, due to the probabilistic nature of LSH, there may not exist a
bucket, which has \textit{all} the documents belonging to an entity. On the other hand, a bucket can have more than one
entity (The number of entities in a bucket is equal to the final number of documents in the set \textit{I$'$}). We call
the entities obtained from each bucket as \textit{partial entities}. So, it is required to combine the results (partial
entities) from all the buckets to get the final resolved view of entities, as discussed below. 

\subsection{Connected Components}\label{subsec:CC}
The results from the buckets need to be combined so that if any two partial entities belonging to different buckets
have a common document, then the entities are combined to get one entity by applying the \textit{Merge} operation on the two
partial entities (which is equivalent to merging all the documents belonging to the two partial entities). The
\textbf{IMM} process ensures that a document cannot belong to more than one partial entity within the
same bucket (because, as soon as a document matches some other document, a new document is created by merging the
original ones and the original ones are removed). Hence, this merging is required for partial entities sharing a common document
but coming from different buckets, rather than for partial entities within the same bucket. 

For the
example in Figure \ref{fig:SampleData}, documents d$_{8}$, d$_{9}$, and d$_{10}$ belonging to the same entity e$_{3}$ do
not share any references to each other. As shown in Figures \ref{fig:FlowAfterDT-LSH1} and \ref{fig:BucketResCC}, documents d$_{8}$ and d$_{9}$ (which has document d$_{11}$ belonging to a different entity e$_{4}$
linked to it through UST) end up in bucket b$_{3}$ based on textual similarity. Similarly, documents d$_{8}$ and
d$_{10}$ end up being in the bucket b$_{4}$. So, there is no such bucket which has all the documents belonging to
e$_{3}$ in it. In bucket b$_{3}$, IMM helps to produce a partial entity e$_{3}^{'}$ = Merge(d$_{8}$, d$_{9}$). In bucket
b$_{4}$, IMM helps to produce another partial entity e$_{3}^{''}$ = Merge(d$_{8}$, d$_{10}$). Now these two partial entities
need to be somehow merged:

We formulate the problem of combining the partial entities from different buckets into the problem of finding the
\textit{connected components} (\textbf{CC}) in an undirected graph. In a distributed setting, connected components
has been used to assign entity ids to matching documents in a similar manner in \cite{distrER}. 

Consider the documents as \textit{nodes} of an
\textit{undirected graph}. There is one node in the graph for each document in the collection D. We create the
\textit{edges} in the graph is as follows: For each partial entity, randomly select one of the nodes from it as a
\textit{central node}. Then, add an edge between the central node and each of the remaining nodes of the partial entity.
This ensures that all the nodes in any partial entity are \textit{connected} to each other (through the central node). 
Consider bucket b$_{1}$ in Figure \ref{fig:FlowAfterDT-LSH1}, which has a partial entity e$_{1}^{'}$ =
\textit{Merge}(d$_{1}$, d$_{2}$, d$_{3}$, d$_{4}$). Selecting d$_{1}$ as the central node for e$_{
1}^{'}$, the edge-list from e$_{1}^{'}$ will be \{d$_{1}$-d$_{2}$, d$_{1}$-d$_{3}$, d$_{1}$-d$_{4}$\}. 

The way in which
the graph is constructed ensures that all the nodes in a partial entity are connected. Also, if any two partial entities
have a node (document) in common, then all the nodes belonging to the two partial entities are connected, and belong to
the same entity. So, one connected component in the graph corresponds to one entity. Hence, by finding the connected
components in the graph constructed as described above, we can combine the results of the buckets. More precisely, the
results from different buckets can be combined to get the resolved entities (see lines \ref{algo:edgeList} to
\ref{algo:mergeCC} in Algorithm \ref{algo:BatchAlgo}) in the following way: (1) Find the connected components in a graph
constructed from the partial entities from all the buckets. (2) Get the final resolved entities by applying the
\textit{Merge} function on the nodes (documents) in each connected component.

For the example in Figure \ref{fig:BucketResCC}, the pair d$_{8}$-d$_{9}$ gets added to the edge-list from bucket b$_{3}$
and the pair d$_{8}$-d$_{10}$ gets added to the edge-list from bucket b$_{4}$. When CC is applied on the edge-list,
d$_{8}$-d$_{9}$-d$_{10}$ emerges as a single connected component c$_{3}$. Then, the Merge function is applied on
d$_{8}$, d$_{9}$, and d$_{10}$ to get the final resolved entity e$_{3}$ = Merge (d$_{8}$, d$_{9}$, d$_{10}$).

Note that the above procedure is functionally equivalent to computing connected components the smaller graph of
partial entities, where an edge denotes two entities sharing some document. However, computing these edges itself
requires traversing and sorting the documents and is therefore not computationally more efficient than the above approach.

\section{Incremental ERLD}\label{sec:IncreMode}
We now show how the ERLD algorithm above can be modified to operate incrementally.
In other words, we describe how a previously resolved entity-document collection can be resolved
with a new set of documents \textit{without} having to re-resolve the entire collection.

First, the inverted-index over the entire
document collection is updated to include the new set of documents. Next, Document Traversal (DT) is performed for each new document.
A single step of DST captures documents reachable `downwards' from each new document, from amongst
the new set as well as the old collection.
Similarly, a single step of UST for can discover references both from the new and old documents. 

Now any old documents obtained through DST or UST are replaced by the corresponding previously resolved entities. 
Thus the traversal-set of a new document can now contain both documents as well as entities: Any documents in
this set are necessarily from the new set of documents, and any entities are those that have been previously resolved.

DT is followed by LSH on the new set of documents. The bucket-ids created by the LSH process on the new set of documents
may contain bucket-ids that were also created earlier LSH was earlier applied on the old documents. The ids of old
documents that got hashed to such bucket-ids are retrieved from the previously created LSH-Index. The corresponding old resolved
entities for these document-ids are retrieved. As a result we now have two types of buckets in the new LSH-index: those that
include old entities (obtained either through traversal or LSH on a new document) and the remaining that contain only new documents. 
The former may lead to updating retrieved old entities in case some new documents need to be merged into them. 

IMM is applied on the new documents as well any entities in each bucket. As in batch mode, partial
entities are generated to be combined using CC. A partial entity in case of Incremental ERLD can be of one of the 3 types: (1) consisting only of new document(s), (2) consisting of new document(s) and old entity/entities, (3) a not updated old entity. If a partial entity is of first or third type, the edge-list is created in the same way as done in ERLD (Section \ref{subsec:CC}). On the other hand, if a partial entity is of type-2, the set of document-ids consists of document-ids for the new documents and the document-ids contained in the old-entities. Edge-list is created for this set of document-ids by assuming one document (node) as a central node and the remaining nodes are connected to it. It is to be noted that, the CC step in Incremental ERLD, happens only on documents (nodes) which are part of the partial entities obtained during Incremental ERLD, and not all the documents (from the old batch and the new set). So, CC happens on the graph corresponding to all the new documents and possibly a few of the old documents. These partial entities consist of old entities and new documents. The
edge-list for the old entities is created in the same way as it is done for partial entities in the batch-mode entity
resolution as described in \ref{sec:batchER}. The CC step may result in updating the entity-ids (and, in turn, the document-entity
map) for some old documents; for example, previously separate entities may get merged because of fresh information
obtained from newly arriving documents. The final resolved entities are obtained by applying the
Merge procedure on the nodes (documents) in each connected component. 

Next the LSH-index is
updated by adding the new buckets (buckets with ids not present in the 
existing LSH-index) as well as adding new documents to previously existing buckets (documents that get mapped to
previously existing bucket-ids). The document-entity map is also updated, which involves updating the entity-ids of some
old documents, as well as adding document-ids and entity-ids for the new documents; in addition, certain previously
resolve entities may disappear by getting merged with others. The updated LSH-index, document-entity
map, and inverted index are re-used for the subsequent incremental ERLD steps, i.e., when fresh sets of documents arrive.

For the example in Figure \ref{fig:SampleData}, assume that document d$_{3}$ was not present when the first batch of
documents was resolved. In the absence of d$_{3}$, it is not possible to get d$_{1}$, d$_{2}$, and d$_{4}$ in the same
bucket. As a result there would be two entities present corresponding to the entity e$_{1}$ in the previously resolved entity-document
collection. One would be e$^{'}_{1}$ = Merge(d$_{1}$, d$_{2}$) (on the basis of traversal), and the other would be
e$^{''}_{1}$ = d$_{4}$. When d$_{3}$ arrives as a part of a new incremental set of documents, it will get linked to
d$_{1}$ and d$_{2}$ (which map to the entity e$^{'}_{1}$) on the basis of DT. Also, since d$_{3}$ has high textual similarity with d$_{4}$, LSH on it will produce at least one bucket-id which has d$_{4}$ (the entity e$^{''}_{1}$) in it. In
this way, d$_{3}$ with its \textit{traversal set} \{e$^{'}_{1}$\} will be present in a bucket with entity e$^{''}_{1}$.
Hence, all documents and previously resolved entities belonging to the desired entity e$_{1}$ land up in the 
same bucket, so the IMM process in this bucket is able to successfully resolve the entiy e$_{1}$. At the same time,
we have avoided re-resolving the entire collection, i.e., the previously resolved entities e$_2$ and e$_3$ are
not even accessed.
\begin{figure}
	\centering
	\includegraphics[scale=0.45,natwidth=1467,natheight=1053]{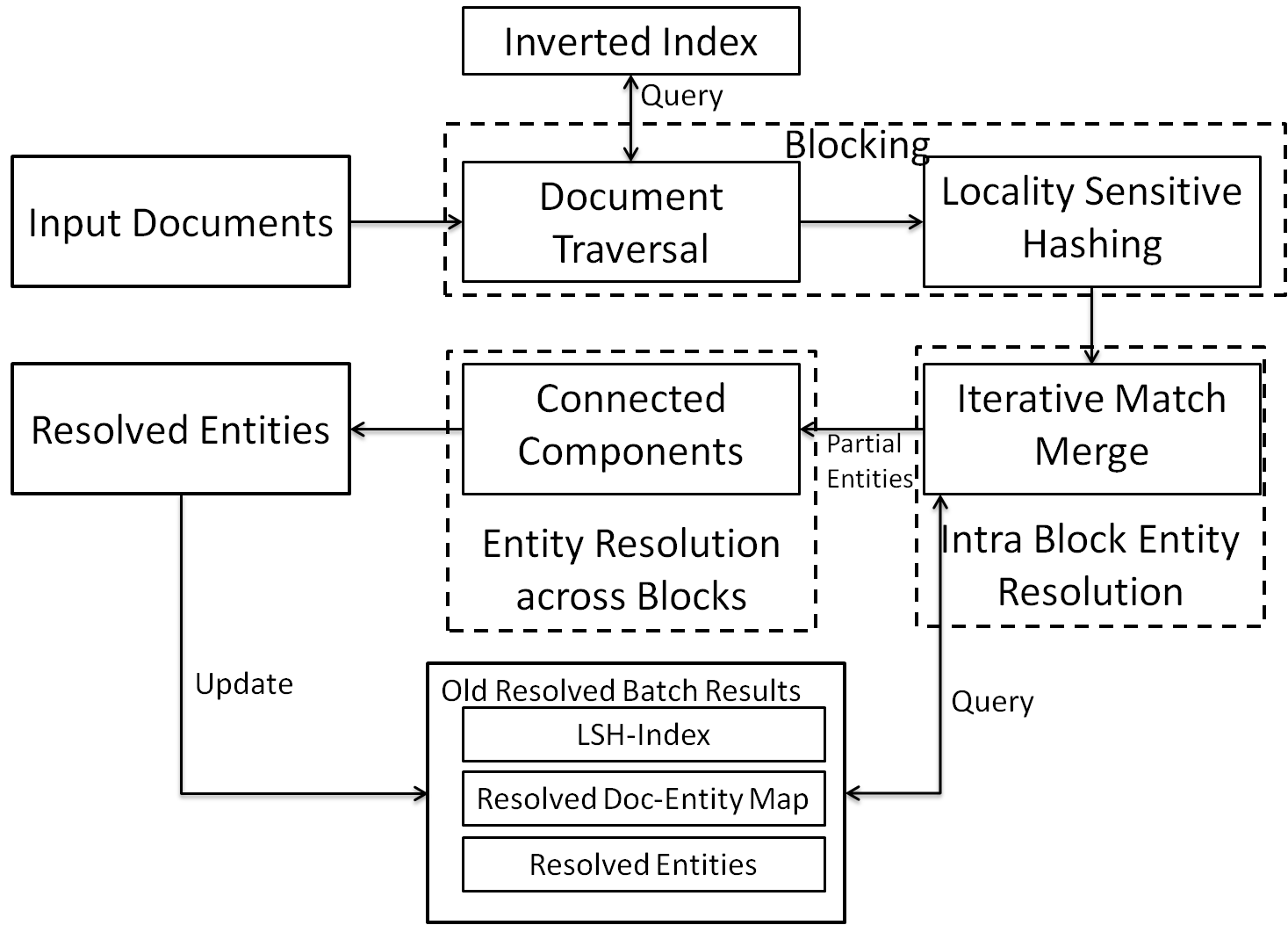}
	\caption{Incremental ERLD}
	\label{fig:IncEntResProcess}
\end{figure}
\section{Experimental Evaluation}\label{sec:exper}
We present performance (execution-time) and quality (precision, recall, and F1-score) results on two data-sets: a
real-world database of companies and a large synthetically generated `residents' database. We first describe the two
data-sets and then present the benchmarks to prove the benefit of i) traversal through enhanced recall, ii) blocking
(based on traversal and LSH) that enables scalability, and iii) incremental resolution that saves execution-time
while maintaining quality as compared to the batch-mode. 

Our implementation was in Java 1.6, and the experiments performed on a Linux (Ubuntu) server with Intel Xeon E7540 2GHz processor, 
64GB RAM and 4 CPUs with 6 Cores each. 

\subsection{Synthetic Residents' Data}\label{subsec:ResidentsData}
We automated the generation of a synthetic `residents' database of documents in the following manner: 
We began with 100,000 seed documents where a seed document
has all the information about an entity. For each such entity, a maximum of 5 documents are created. These documents
belong to one of the following 5 domains: Voter-Card, PAN Card, Driving-Licence, Bank Account Details, and Phone
Connection Application. An entity can have a maximum of 1 document per domain. We control the creation of a document of
a particular type for an entity using a random-number generator. To create a document from the seed document, values of
some of the attributes of the person are retained. The values of the attributes of an entity across documents are varied
by using different methods. For example, the address of the different documents for the same entity need not be same.
Even if the address is same, a slight variation is inserted by skipping a few characters to introduce typographical
errors. The first name of person is varied in some cases by omitting a few 
characters, swapping the first and last name, or omitting the middle name. Also, some of the attributes are not given
any value. For example, e-mail id of a person may be present in 2 of his documents and absent in others. 

Once the documents for an entity are
created, documents are also created for some related entities, such as parent, child, spouse, neighbour, etc.
This also adds ambiguity to the resolution of entities as documents of related entities have a considerably high textual
similarity. Once all the documents for a person are generated, references are added between them in a random manner. The
documents belonging to the same person share explicit references between them. 

References between documents belonging
to different entities are not present in this data-set. Note that such references would need to be treated differently from 
references between documents belonging to the same entity. In most cases document structure would make it easy
to distinguish \textit{explicit} references that point to documents of another entity, e.g., a field labeled `father's passport number'.
Sometimes however, and certainly for \textit{implicit} references (i.e., in textual description fields), such a distinction 
may not be possible. We do not address (a) disambiguation of such references, nor do we exploit (b) the potential of
collective entity resolution by using relationships between entities to improve performance, as in \cite{p:indrajit:ColBatch}. 

We generated 580,363 documents corresponding to 209,501 entities. 
The average traversal set size for a document is 1.59 with a range from 0 to 4. 
Out of these, the traversal set size for 158,708 documents is 0, for 142,770 documents it is 1, for 124,873 documents it is 2, for 87,374 documents it is 3, and for the remaining 66,638 documents it is 4. 

\textit{Benefit of Document References:} Table \ref{tab:ResiTravBenefit} demonstrates the additional benefit of using  inter-document references 
via DT during blocking of documents:  Consider add the rule
the rule \emph{sameName}(\textit{r},\textit{r'}) \texttt{AND} \emph{sameAddress}(\textit{r},\textit{r'}) \texttt{AND}
(\emph{isInTraversalSet}(\textit{r},\textit{r'}) \texttt{OR} \emph{isInTraversalSet}(\textit{r'},\textit{r})),
which includes clauses testing for inter-document references. We witness a 2\% increase in recall without any loss in precision
when this rule exploiting inter-document references is used, as compared to the results without such a rule.

\begin{table}
\centering
		\begin{tabular}{|p{2.4cm}|c|c|c|p{1.2cm}|}
		\hline
		\textbf{Algorithm} & \textbf{Precision} & \textbf{Recall} &	\textbf{F1 Score} \\ \hline
		References not used &	1.00	& 0.96	& 0.98 \\ \hline
		References used &	1.00 &	0.98	& 0.99 \\		
		\hline
		\end{tabular}
		\caption{Benefit of Traversal in Residents' Data Batch-mode ERLD. Number of documents = 580,363}
		\label{tab:ResiTravBenefit}
\end{table}
\begin{figure}
\centering
	\centerline{\includegraphics[scale=0.5,natwidth=1256,natheight=453]{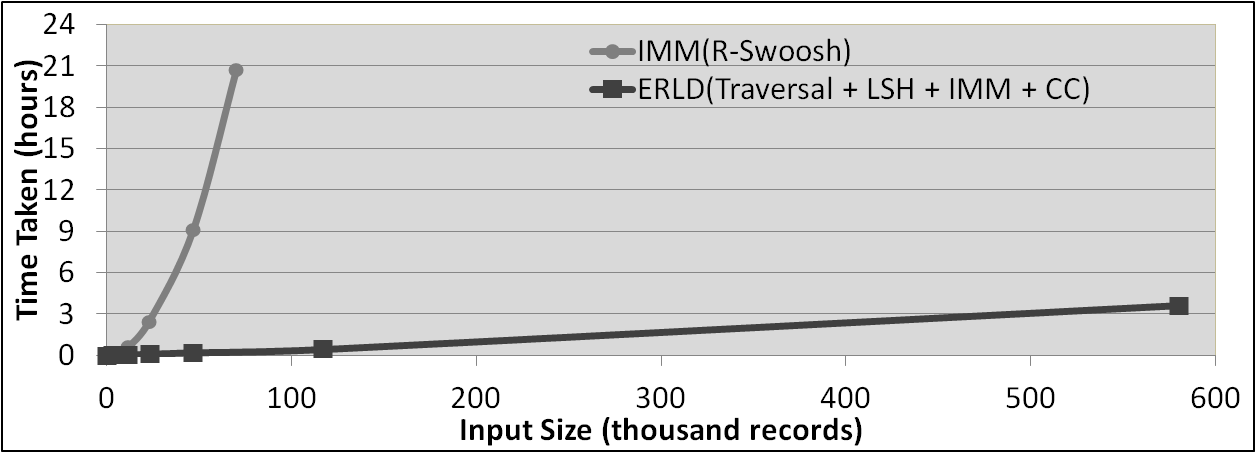}}
	\caption{Scalability Analysis for Residents' Data}
	\label{fig:ResidentsScalability}
\end{figure}
\begin{table}
\centering
		\begin{tabular}{|p{4.6cm}|c|c|c|c|}
		\hline
		 & \textbf{Precision} & \textbf{Recall} & \textbf{F1 Score} &	\textbf{Time Taken (hh:mm)} \\ \hline
		ERLD(X+Y) &	1.00	& 0.98	& 0.99 &	03:38 \\ \hline
		ERLD(X) &	1.00 &	0.91	& 0.95 &	03:29 \\ \hline
		IncrementalERLD(Y) &	1.00 &	0.98 &	0.99 &	01:20 \\ \hline
		ERLD(X)+ERLD(X+Y) &	1.00 &	0.98 &	0.99 &	07:07 \\ \hline
		ERLD(X)+IncrementalERLD(Y) &	1.00 &	0.98 &	0.99 &	04:49 \\ 
		\hline
		\end{tabular}
		\caption{Incremental-mode ERLD for Residents' Data, documents resolved using Batch-mode:X=551,249,
documents resolved using Incremental-mode:Y=29,114, Total number of documents (X+Y)=580,363}
		\label{tab:ResiDataInc}
\end{table}

\textit{Performance gains from DT+LSH-based blocking:} Blocking based on traversal and LSH has a much better execution-time performance (refer Figure
\ref{fig:ResidentsScalability}) than IMM without any blocking. Whereas the Iterative Match-Merge (IMM) based approach takes about 21 hours to resolve
70,291 documents, our batch-mode entity resolution process takes less than 4 hours to resolve 580,363 documents without
any change in the quality of results (both having precision = 1.00, recall = 0.95 for 70,291 documents).

\textit{Efficiency and accuracy of Incremental ERLD:} The 580,363 documents were randomly distributed into 2 sets X and Y, such 
that X had roughly 95\% (551,249) of the documents, and Y
had the remaining 5\% of the documents (29,114). From Table \ref{tab:ResiDataInc} it can be observed that first performing 
ERLD on a data-set(X) and then Incremental ERLD on the new batch of data(Y), results in saving of more than 2 hrs with 
respect to performing ERLD twice first on initial data-set (X) and then on combined data-set(X+Y), without
any compromise in quality of results.

\subsection{Companies' Data}\label{subsec:CompaniesData}
This real-life data-set contained information about companies with attributes such as, 
Company Name, Email, Contact Phone, Company Fax, Address Line 1, Address Line 2, City, State, Postal Code, etc. 
There were a total of  669,245 documents; out of these 2196 pairs were manually labeled by domain experts,
resulting in 1697 unique entities. A union of the 2196 pairs of documents gave us 3801 distinct documents (since some
a documents were present in more that one pair).

For this data set we used a supervised SVM-based binary classifier (similar to \cite{stringSimMetrics}) as the \textit{Match} function (instead of rules).
for each pair of documents, features were computed as follows:
Eight document attributes were used. The values of an attribute from two
documents in a pair was compared using 7 string similarity metrics \{Overlap Coefficient, Jaro-Winkler, SoftTfIdf, Soundex, Monge-Elkan, Jaccard Similarity, Cosine Similarity\} each giving a score in the range of 0 to 1. 
This resulted in a $56 (=8$x$7)$ dimensional feature-vector for each pair of documents. 
We used 70\% of the matching pairs (2196) as positive samples for the training set. All the document pairs other than those present in
the labeled matching pairs $(^{3801} C_2 - 2196)$ are assumed to be non-matching. Only, 3\% of such pairs were taken 
as negative samples for training. An SVM with linear kernel function and cost factor $C = 0.06$ was used.

\begin{table}
\centering
		\begin{tabular}{|p{3.0cm}|c|c|c|p{1.2cm}|}
		\hline
		\textbf{Algorithm} & \textbf{Precision} & \textbf{Recall} &	\textbf{F1 Score} \\ \hline
		ERLD(X+Y) &	0.974	& 0.878	& 0.924 \\ \hline
		ERLD(X) &	0.973	& 0.800	& 0.878 \\ \hline
		IncrementalERLD(Y) &	0.974 &	0.876	& 0.922 \\		
		\hline
		\end{tabular}
		\caption{Incremental-mode ERLD for Companies' Data. X=3605, Y=196, Total number of documents(X+Y)=3801}
		\label{tab:CompaniesQualityInc}
\end{table}
\begin{table}
\centering
		\begin{tabular}{|p{2.5cm}|c|c|c|c|}
		\hline
		\textbf{Algorithm} & \textbf{Precision} & \textbf{Recall} &	\textbf{F1 Score} \\ \hline
		AllPairs+CC &	0.971	& 0.885	& 0.926 \\ \hline
		LSH+IMM+CC &	0.974 &	0.878	& 0.924 \\		
		\hline
		\end{tabular}
		\caption{Quality comparison on real-world Companies' Data: LSH+IMM+CC vs. AllPairs+CC. Number of
documents = 3801}
		\label{tab:CompQualityBatch}
\end{table}
\begin{figure}
	\centering
	\centerline{\includegraphics[scale=0.5,natwidth=1144,natheight=669]{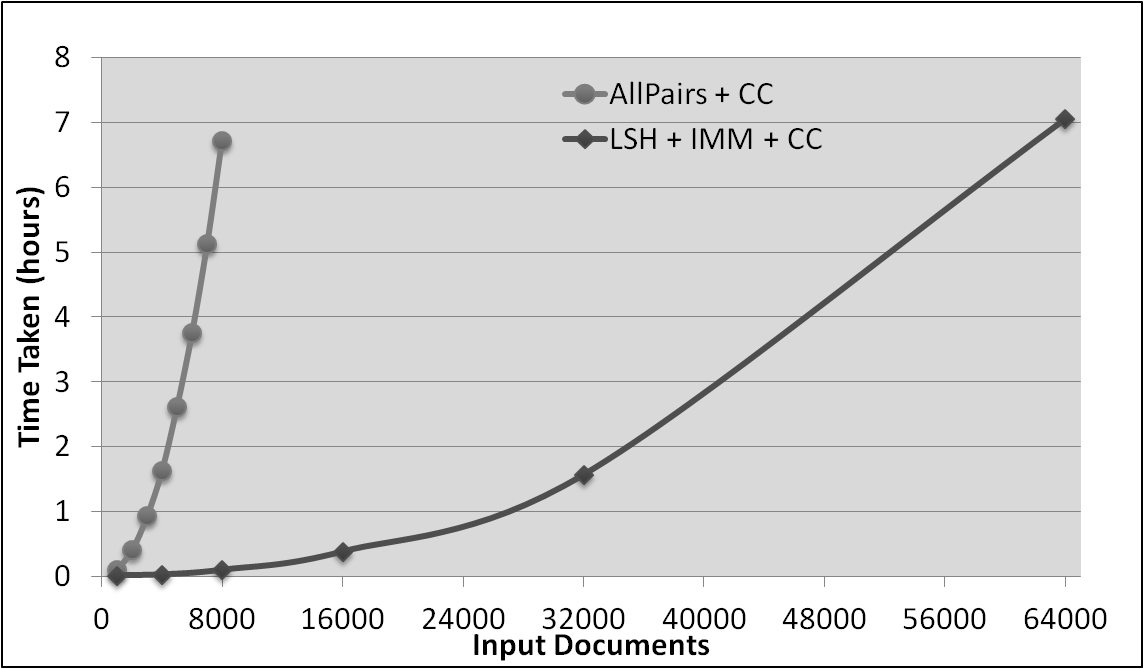}}
	\caption{Scalability Analysis for Companies' Data}
	\label{fig:CompScal}
\end{figure}

\textit{ERLD vs all-pairs comparison:} Our ERLD algorithm based on blocking rather than
exhaustive all-pairs comparison using a Match function, which in this case was our machine-learned classifier.
We observe precision-recall and time-taken of two approaches, viz., (i) Compare each document with every other document. 
Create an edge for every matching pair and apply connected components on such a graph to combine all the documents which are part of same entity.
(ii) ERLD, i.e., LSH followed by Iterative Match-Merge followed by Connected Components to resolve entities. (Note: there is no DT stage
since this data set does not contain references).

Figure \ref{fig:CompScal} shows that the time taken by the second approach is considerably less than the time taken by
the first approach and Table \ref{tab:CompQualityBatch} shows that this is achieved without any significant loss in the
quality of the results. 

\textit{Incremental vs Batch on Companies' data:} 
Nevertheless, as depicted in Table \ref{tab:CompaniesQualityInc}, incrementally resolving the set Y (196
documents) against previously resolved set X (3605 documents) gives the same precision-recall as obtained by resolving
all the documents (X+Y) together. 
Of course, the performance benefits of incremental vs batch resolution were less significant (3 minutes for Y vs 4 minutes for X+Y)
than earlier, as the Companies' data set is considerably smaller than the synthetic residents data.
\section{Related Work}\label{sec:relWork}
Blocking for entity resolution has been often used \cite{distrER}.
Various iterative and non-iterative multiple blocking techniques for entity resolution have been discussed in detail 
in \cite{swooshThesis}, such as Locality Sensitive Hashing, Q-gram Based Indexing etc. 
Locality Sensitive Hashing (LSH) based data-blocking for Iterative Record Linkage has been discussed
in \cite{harra}, which
iteratively creates hash tables for records based on minhash functions and resolves the records being hashed to the same
key. The resolved records at the end of each iteration are used as input records for next iteration. Various indexing
(blocking) techniques for Entity Resolution like Sorted Neighborhood Indexing and Q-gram Based Indexing have been
discussed in \cite{indexingSurvey}. A set of blocking schemes that work for heterogeneous
schema were proposed in \cite{p:beyond100m}.
In \cite{p:pay-as-u-go} ``hints" are used after blocking, where the records that are more likely to get merged 
are compared before others, so that the partial results of Entity Resolution are available as they are generated.
Our ERLD approach additionally includes inter-document references
to enhance the benefits and quality of blocking.

We highlight the importance incremental-mode entity resolution where a small batch of documents is resolved, against an already
resolved entity collection, and propose an efficient solution. However, the concept of incremental processing is
not new and has been proposed in \cite{p:DataHarmo}. In the domain of entity resolution, incremental duplicate detection has been
attempted in \cite{efficientIncr}. However they process only one additional document at a time, while we process a batch 
of newly arrived documents and we perform complete entity resolution. Resolution of small set of entities has been 
attempted in \cite{realtimeER,queryIndra} for query interface and not in incremental manner, after bulk entity resolution has already taken place.

We have considered inter-document references that indicate that two documents belong to the same entity. However, such
references could also indicate inter-entity relationships. In such cases our approach can possibly be used in as a precursor
to collective entity resolution \cite{p:indrajit:ColBatch}.

Our algorithm is based on traversal of the document
graph. Graph traversal has been used in many other related domains, such as, `keyword search on graph data' 
\cite{p:banks1, p:sagiv1}. Most of these approaches first convert the data into a graph, and then use various techniques
for bi-directional graph traversal to discover a node that is reachable from the starting nodes. In our case, however,
we never explicitly instantiate the entire document-graph; instead we only work on a small local neighbourhood
of each document obtained using direct references (DT) and an inverted index (UST). 
Further, keyword-search algorithms
include intermediate nodes in paths discovered between keyword nodes, resulting in a Steiner Tree problem.
However, we assume that inter-document references are direct, rather than indirect via intermediate nodes.

\section{Conclusions}\label{sec:conc}
We have shown how inter-document references can be leveraged to enhance the quality of entity resolution results through graph-traversal. We have then proposed a graph-traversal and Locality Sensitive Hashing based blocking scheme for a scalable solution to the Entity Resolution problem without any significant loss in the quality of results. We have further shown how to incrementally resolve a new set of documents against a pre-resolved entity-document collection without having to re-resolve all the documents.

\bibliographystyle{ieeetr}
\end{document}